# Anomalous Hall effect in *L1*₀-MnAl films with controllable orbital two-channel Kondo effect

L. J. Zhu*, S. H. Nie and J. H. Zhao†

*State Key Laboratory of Superlattices and Microstructures, Institute of Semiconductors, Chinese Academy of Sciences, P. O. Box 912, Beijing 100083, China*

**Abstract：** The anomalous Hall effect (AHE) in strongly disordered magnetic systems has been buried in persistent confusion despite its long history. We report the AHE in perpendicularly magnetized $L1_0$-MnAl epitaxial films with variable orbital two-channel Kondo (2CK) effect arising from the strong coupling of conduction electrons and the structural disorders of two-level systems. The AHE is observed to excellently scale with $\rho_{AH}/f = a_0\rho_{xx0} + b\rho_{xx}^2$ at high temperatures where phonon scattering prevails. In contrast, significant deviation occurs at low temperatures where the orbital 2CK effect becomes important, suggesting a negative AHE contribution. The deviation of the scaling agrees with the orbital 2CK effect in the breakdown temperatures and deviation magnitudes.

* zhulijun0@gmail.com
† jhzhao@red.semi.ac.cn

The anomalous Hall effect (AHE) is the most prominent phenomena existing in magnetic materials. The AHE has received renewed interests in recent years due to its rich phenomenology that defies the standard classification methodology and prompting conflicting reports that claim the dominance of various microscopic processes [1-3]. The scaling between longitudinal resistivity ($\rho_{xx}$) and anomalous Hall resistivity ($\rho_{AH}$) can provide access to the detailed mechanisms of the AHE. Present theories predict a scaling of $\rho_{AH} \sim \rho_{xx}^n$, with $n = 2$ for side jump and intrinsic contribution [4,5], and $n = 1$ for skew scattering [6]. The intrinsic contribution is related to the topological Berry curvature of band structure [4], while the extrinsic side jump and skew scattering result from spin-orbit interaction-induced asymmetric impurity scattering of conduction electrons [5,6]. Notably, in contrast to the conventional picture that $\rho_{AH}$ scales with the total resistivity irrespective of its sources [3], recent experimental studies have revealed that both $\rho_{AH}$ and the scaling relation are qualitatively different for various types of electron scattering. Phonon scattering was found to have no distinguishable contribution to extrinsic part of $\rho_{AH}$ in ferromagnetic Fe, Co, Ni, $L1_0$-Mn$_{1.5}$Ga films, Co/Pt multilayers, and paramagnetic Ni$_{34}$Cu$_{66}$ films [7-12]. When $\rho_{xx}$ is dominated by phonon and static defect scattering, the AHE has a scaling of $\rho_{AH} = a_0\rho_{xx0} + b\rho_{xx}^2$, where $\rho_{xx0}$ is the residual resistivity, $a_0\rho_{xx0}$ is the extrinsic contribution from both side jump and skew scattering, and $b$ is the intrinsic anomalous Hall conductivity (AHC). In magnetic materials with relatively low Curie temperature, the temperature ($T$) induced variation of the magnetization must be taken into account [13]. For the simplest case that $\rho_{AH}$ scales proportionally with magnetization ($M$) which is not the fact in some magnetic systems [14,15], the AHE scaling can be modified into a more general expression explicitly as $\rho_{AH}/f = a_0\rho_{xx0} + b\rho_{xx}^2$. Here, $f = M / M_0$, where $M_0$ is magnetization at 0 K; the $T$-independence of $a_0$ and $b$ is not considered. In the presence of strong disorder effects (dirty metal [1]), the AHE scaling is more complex. In the hopping conduction regime, the AHE was reported to scale as $\rho_{AH} \sim \rho_{xx}^{0.4}$ in

Fe$_{3-x}$Zn$_x$O$_4$ and (Ga,Mn)As [16,17], $\rho_{AH} \sim \rho_{xx}^{0.5 \sim 0.3}$ in Ti$_{1-x}$Co$_x$O$_{2-\delta}$ [18], and $\rho_{AH} \sim \rho_{xx}^0$ in polycrystalline FePt [19], respectively. It has remained an open question whether or not weak localization and electron-electron interaction contribute to $\rho_{AH}$ [19-21]. Therefore, further experimental and theoretical attempts are crucial for better understanding of the AHE, especially in disordered magnetic systems.

The Kondo effect is a striking consequence of conduction electrons coupling with localized spin or "pseudospin" impurities. The one-channel Kondo systems are normally found to have an important AHE contribution arising from the coupling of Kondo centers and conduction electrons [22,23]. However, the interplay between the AHE and the two-channel Kondo (2CK) effect displaying exotic non-Fermi-liquid (NFL) physics has remained elusive. Aliev *et al.* predicted an exact cancellation of the AHE contributions from two channels of quadruple 2CK effect when the magnetic field is not so large to eliminate the channel symmetry and presented an experimental evidence in U$_{0.9}$Th$_{0.1}$Be$_{13}$ [24]. In contrast, Sato *et al.* found a negative AHE contribution and a breakdown of the scaling $\rho_{AH} \sim \rho_{xx}^2$ at low $T$ in CeRu$_2$Si$_2$ single crystal exhibiting NFL behaviors [25]. To our best knowledge, there have been no reports about the AHE for orbital 2CK effect arising from coherent tunneling scattering of localized "pseudo-spin" 1/2 impurities between two independent quantum wells, i.e. dynamical disorder of structural two-level-systems (TLSs) [26]. The orbital 2CK effect has a magnetic field ($H$)-independent contribution to $\rho_{xx}$ which scales with ln$T$ (Kondo scattering), $T^{1/2}$ (NFL regime), and a deviation from $T^{1/2}$ (breakdown of NFL physics) in three characteristic $T$ regimes, respectively.

Most recently, we observed robust and controllable orbital 2CK effect evidenced by the three-regime resistivity increase at low temperatures in $L1_0$-MnAl ferromagnetic films [27], which may provide an experimental access to the AHE in the presence of the orbital 2CK effect. On the other hand, $L1_0$-MnAl films are ideal for the AHE studies due to their giant perpendicular magnetic anisotropy ($\sim 10^6$ erg/cc), well-defined perpendicular hysteresis loops, and large values of $\rho_{AH}$ (e.g. 2~7.5 $\mu\Omega$ cm in $Mn_{1.5}Al$ films grown on AlAs-buffered GaAs) [28-32]. However, there has been no report about the temperature dependence of $\rho_{AH}$ or the scaling of the AHE in MnAl to date. In this letter, we present the AHE in $L1_0$-MnAl epitaxial films with variable orbital 2CK effect. The AHE scaling well agrees with $\rho_{AH}/f = a_0\rho_{xx0} + b\rho_{xx}^2$ in high-$T$ regime where phonon scattering is dominating but seriously breaks down at low temperatures most likely due to a negative AHE contribution of orbital 2CK physics.

A series of 30 nm-thick $L1_0$-MnAl films with Mn/Al atom ratio of 1.1 were epitaxially grown on semi-insulating GaAs (001) substrates by molecular-beam epitaxy at 200, 250, 300, 350 and 400 ℃, respectively, and then capped with a 4 nm-thick $Al_2O_3$ layer to prevent oxidation [27,30]. The good homogeneity and sharp interface of these films are confirmed by the cross-sectional high-resolution transmission electron microscopy images [29]. More details of structures, magnetic and transport properties could be found elsewhere [27-29]. The values of $f$ of these films were determined by a Quantum Design SQUID-5 system. The films were patterned into 60 $\mu$m-wide Hall bars with adjacent electrode distance of 200 $\mu$m using photolithography and ion-beam etching for transport measurements (see Fig. 1(a)). Both Hall resistivity $\rho_{xy}$ and $\rho_{xx}$ were measured as a function of $T$ (2-300 K) and perpendicular $H$ by a Quantum Design PPMS-9 system.

Figure 1(b) shows an example of $\rho_{xx}$-$H$ curves at different $T$ for $L1_0$-MnAl films ($T_s = 200$ ℃). For all the temperatures concerned in the work, $\rho_{xx}$ varies little with $H$ in comparison to the large value of itself, which is further confirmed by the small magnetoresistance (<0.5%) [27], suggesting a negligible spin fluctuation and spin waves in this material probably due to their strong anisotropic field [33]. These features hold for all the films with different $T_s$. Figure 1(c) summarizes $\rho_{xx}$ at $H = 0$ T for all the $L1_0$-MnAl films as a function of $T$. Each film shows a resistivity minimum at characteristic temperature ($T_m$), beyond which $\rho_{xx}$ increases linearly with increasing $T$ due to the dominating phonon scattering. As $T$ decreases from $T_m$ towards 0 K, $\rho_{xx}$ shows an intriguing increase arising from the orbital 2CK effect [27]. More detailed studies show that the values of $T_m$ are about 190, 41, 18, 19, and 24 K for $T_s = 200$, 250, 300, 350, and 400 ℃, respectively [27]. By subtracting the contribution from the orbital 2CK effect, the residual resistivity $\rho_{xx0}$ (i.e. the offsets $\rho_1 \approx \rho_2$ in ref. 27) was determined to be 218.5, 174.7, 132.2, 135.7, and 141.1 $\mu\Omega$ cm for the films grown at 200, 250, 300, 350 and 400 ℃, respectively.

In order to investigate the AHE and possible interplay with the 2CK effect, we first evaluated $\rho_{AH}$ in $L1_0$-MnAl films by subtracting the ordinary Hall component $R_0 H$ (determined from a linear fit to the high-field regions of the $\rho_{xy}$-$H$ curves up to $\pm 7$ T) following the empirical relationship of $\rho_{xy} = R_0 H + R_s M$ with $\rho_{AH} = R_s M$. Figure 2(a) shows $\rho_{AH}$-$H$ curves measured at 300 K for all these films with different $T_s$. $\rho_{AH}$ and switching field of $\rho_{AH}$-$H$ curves show significant variation among these samples, which agrees with the thermally-tuned structural disorders and magnetic properties [30], Kondo temperature, and the TLS density in these films [27]. The film grown at 200 ℃ shows the most gradient hysteresis, coinciding with the observation of the strongest orbital Kondo effects among these studied films. Figure 2(b) shows the $T$ profile of $\rho_{AH}$ in these films. $\rho_{AH}$ varies between 1.28 $\mu\Omega$ cm and 1.93 $\mu\Omega$ cm, which is smaller than previously reported 2~7.3 $\mu\Omega$ cm in $Mn_{1.5}Al$ films [31,32]. In common magnetic metals, e.g. Fe and $L1_0$-$Mn_{1.5}Ga$ [7,10], $\rho_{AH}$ are always observed to decrease monotonically as $T$ decreases. In striking contrast, $\rho_{AH}$ of our $L1_0$-MnAl films shows non-monotonic $T$ dependence. For $T_s \geq 300$ ℃, as $T$ decreases from 300 K to 2 K, $\rho_{AH}$ first drops monotonically, reaches the minimum around 50 K, and finally climbs up, which is very similar with that in strongly disordered FePt polycrystalline films with hopping conduction [20]. For $T_s = 200$ ℃ and 250 ℃, $\rho_{AH}$ even increases monotonically upon cooling down. So far, there have been few report on similar anomaly in $T$-dependence of $\rho_{AH}$ [14]. As is shown in Fig. 3(a), with increasing $T$ from 2 K to 300 K, $f$ (here $M_0$ takes the value at 2 K) changes little for the films grown at beyond 300 ℃, whereas reduces by 40% and 50% for the films grown at 250 ℃ and 200 ℃, suggesting a remarkable reduction of Curie temperature. The slightly faster increase in $M$ at lowest temperatures for $L1_0$-MnAl with $T_s$ =200 ℃ could be due to possible compensating alignment of Mn atoms at different lattice sites as was widely discussed in chemically disordered MnGa [34,35]. Notably, the corrected anomalous Hall resistivity, $\rho_{AH}/f$, decreases monotonically with decreasing $T$ (see Fig. 3(b)). However, we infer that the occurrence of orbital 2CK effect at low $T$ could be another reason for such a complex temperature dependence of $\rho_{AH}$ in addition to the $T$-dependent $M$, as is further discussed by the scaling behaviors.

To distinguish the different contributions of $\rho_{AH}$, we checked the scaling behavior of AHE in the $L1_0$-MnAl in Fig. 3(c). $\rho_{AH}/f$ shows a good linear correlation with $\rho_{xx}^2$ in the high $T$ regime ($T > T_m$) for all these films with different $T_s$, consistent with scaling $\rho_{AH}/f = a_0\rho_{xx0} + b\rho_{xx}^2$. The derived values of $a_0$ and $b$ are summarized in Fig. 4(a) as a function of $T_s$. $a_0$ changes from large negative to vanishingly small positive values as $T_s$ increases from 200 ℃ to 400 ℃, which should be due to the reducing extrinsic AHE contribution and static defects, consistent with the observation in $L1_0$-FePt films [13]. $b$ decreases from ~$1.8 \times 10^{-3}$ $\mu\Omega^{-1}$ cm$^{-1}$ at $T_s = 200$ ℃ to $6.0 \times 10^{-5}$ $\mu\Omega^{-1}$ cm$^{-1}$ at high $T_s$. In the MnAl films with low $T_s$, the opposite signs of $a_0$ and $b$ indicate that the extrinsic skew scattering and side-jump contribute to AHC in an opposite way to that of intrinsic contribution as observed in FePt and Fe films [7,13]. The strong $T_s$ dependence of intrinsic AHC $b$ should be attributed partly to the different $T$ dependence of $M$ (see Fig. 3(a)) and partly to the modification of Fermi surface by structural disorders (especially the chemical disorder) [10,13]. As shown in Fig. 4(b), the carrier density $p$ estimated from ordinary Hall coefficient $R_0$ following the single-band model varies remarkably among the MnAl films with different $T_s$, confirming the significant variation of the Fermi surface properties with structural disorders. It should be pointed out that the observed variation of $b$ in the samples with the same composition but different degree of structural disorders reminds us that one should take care when trying to establish the AHE scaling by assuming a constant $b$ or $n$ in a group of samples with different composition.

More intriguingly, the experimental data significantly deviate from the expectation of $\rho_{AH}/f = a_0\rho_{xx0} + b\rho_{xx}^2$ below 150 K and 25 K for $T_S$=200 ºC and 250 ºC, respectively. For other three samples with $T_m < 25$ K, the breakdown of the scaling law is not evident. The excellent agreement between deviation temperatures of the scaling and $T_m$ below which orbital 2CK effect become important is highly reminiscent of a close relation between this breakdown and the orbital 2CK effect. Importantly, the magnitudes of the deviation at 2 K for different samples are also consistent with their strength ($T_K$, $\alpha$, and $\beta$ in Ref. 27) of the orbital 2CK effect. In fact, such breakdown of the scaling can be readily understood because the 2CK effect gives a positive contribution to $\rho_{xx}$ but a negative or zero contribution to $\rho_{AH}$ [24,25]. Here, there seems a negative contribution to $\rho_{AH}$ of our $L1_0$-MnAl films below $T_m$, which is consistent with the observation in quadruple 2CK system of CeRu$_2$Si$_2$ single crystal [25]. Unfortunately, the AHE for the 2CK effects is difficult to treat theoretically at present. More theoretical and experimental efforts are needed in the future for better understanding these intriguing AHE scaling behaviors. It should be mentioned that other scaling laws in literature which based on picture that $\rho_{AH}$ scales with the total resistivity irrespective of its sources cannot solve above breakdown problem [36].

In conclusion, we have systematically presented the AHE in ferromagnetic $L1_0$-MnAl epitaxial films with variable orbital two-channel Kondo effect. The AHE scaling is observed to follow $\rho_{AH}/f = a_0\rho_{xx0}+b\rho_{xx}^2$ at high temperatures where phonon scattering prevails. However, the AHE significantly deviates from it at low temperatures where the orbital 2CK effect becomes important. The breakdown of the scaling seems closely correlated to the orbital 2CK effect which contributes to the longitudinal and anomalous Hall resistivity in different ways. A remarkable tuning of the intrinsic AHC and the carrier density is also observed by varying structural disorders, indicating a significant modification of the Fermi surface properties. These results put forwards new inspiring questions for future theoretical and experimental studies of the AHE.

We acknowledge X. F. Jin for discussions. This work was partly supported by NSFC (Grand No. 61334006) and MOST of China (Grant No. 2015CB921503).

[1] N. Nagaosa, J. Sinova, S. Onoda, A. H. MacDonald, and N. P. Ong, *Rev. Mod. Phys.* **82**, 1539 (2010).

[2] D. Xiao, M.C. Chang, and Q. Niu, *Rev. Mod. Phys.* **82**, 1959 (2010).

[3] P. He, L. Ma, Z. Shi, G. Y. Guo, J. G. Zheng, Y. Xin, and S. M. Zhou, *Phys. Rev. Lett.* **109**, 066402 (2012).

[4] R. Karplus and J. M. Luttinger, *Phys. Rev.* **95**, 1154 (1954).

[5] Berger, *Phys. Rev. B* **2**, 4559 (1970).

[6] J. Smit, Physica (Amsterdam) **21**, 877 (1955).

[7] Y. Tian, L. Ye, and X. F. Jin, *Phys. Rev. Lett.* **103**, 087206 (2010).

[8] D. Z. Hou, Y. F. Li, D. H. Wei, D. Tian, L. Wu, and X. F. Jin, *J. Phys.: Condens. Matter.* **24**, 482001 (2012).

[9] L. Ye, Y. Tian, X. Jin, and D. Xiao, *Phys. Rev. B* **85**, 220403 (2012).

[10] L. J. Zhu, D. Pan, and J. H. Zhao, *Phys. Rev. B* **89**, 220406 (R) (2014)

[11] F. Zhang, F. S. Wen, Y. F. Lu, W. Li, Y. F. Lu, Z. Y. Liu, B. Xu, D. L. Yu, J. L. He, and Y. J. Tian, *J. Appl. Phys.* **110**, 033921 (2011).

[12] Y. Li, D. Hou, G. Su, L. Ye, Y. Tian, J. Xu, G. Su, and X. Jin, *Europhys. Lett.* **110**, 27002 (2015).

[13] M. Chen, Z. Shi, W. J. Xu, X. X. Zhang, J. Du, and S. M. Zhou, *Appl. Phys. Lett.* **98**, 082503 (2011).

[14] R. Mathieu, A. Asamitsu, H. Yamada, K. S. Takahashi, M. Kawasaki, Z. Fang, N. Nagaosa, and Y. Tokura, *Phys. Rev. Lett.* **93**, 016602 (2004).

[15] A. Fert, A. Friederich, and A. Hamzic, *J. Magn. Magn. Mater.* **24**, 231 (1981).

[16] D. Venkateshvaran, W. Kaiser, A. Boger, M. Althammer, M. S. R. Rao, S. T. B. Goennenwein, M. Opel, and R. Gross, *Phys. Rev. B* **78**, 092405 (2008).

[17] D. Chiba, A. Werpachowska, M. Endo, Y. Nishitani, F. Matsukura, T. Dietl, and H. Ohno, *Phys. Rev. Lett.* **104**, 106601 (2010).

[18] H. Toyosaki, T. Fukumura, Y. Yamada, K. Nakajima, T. Chikyow, T. Hasegawa, H. Koinuma, and M. Kawasaki, *Nature Mater.*, **3**,221 ( 2004).

[19] Y. M. Xiong, P. W. Adams, and G. Catelani, *Phys. Rev. Lett.* **104**,076806 (2010).

[20] A. Langenfeld and P. Wölfle, *Phys. Rev. Lett.* **67**,739 (1991).

[21] L. M. Lu, J. W. Cai, Z. Guo, and X. X. Zhang, *Phys. Rev. B* **87**, 094405 (2013).

[22] G. R. Stewart, Z. Fisk, and J. O. Willis, *Phys. Rev. B* **28**, 171 (1983).

[23] P. Coleman, P. W. Anderson, and T. V. Ramakrishnan, *Phys. Rev. Lett.* **55**, 414 (1985).

[24] F. G. Aliev, H. E. Mfarrej, S. Vieira, and R. Villar, *Europhys. Lett.* **34**, 605-610 (1996).

[25] H. Sato, Y. Aoki, J. Urakawa, H. Sugawara, Y. Onuki, T. Fukuhara, and K. Maezawa, *Phys. Rev. B* **58**, R2933 (R) (1998).

[26] D. L. Cox and A. Zawadowki, *Adv. Phys.* **47**, 599-942 (1998).

[27] L. J. Zhu, S. H. Nie, P. Xiong, P. Schlottmann, and J. H. Zhao, *Nat. Commun.* (Accepted); Preprint at arXiv: 1506.04360(2015).

[28] L. J. Zhu, S. H. Nie, and J. H. Zhao, *Chin. Phys. B* **22**,118505 (2013).

[29] S. H. Nie, L. J. Zhu, J. Lu, D. Pan, H. L. Wang, X. Z. Yu, J. X. Xiao and J. H. Zhao, *Appl. Phys. Lett* **102**, 152405 (2013).

[30] S. H. Nie, L. J. Zhu, D. Pan, J. Lu and J. H. Zhao, *Acta Phys. Sin.* **62**, 178103 (2013).

[31] J. D. Boeck, T. Sands, J. P. Harbison, A. Scherer, H. Gilchrist, T. L. Cheeks, M. Tanaka and V. G. Keramidas, *Electronics Lett.* **29**, 421 (1993).

[32] M. L. Leadbeater, S. J. Allen, F. DeRosa, J. P. Harbison, T. Sands, R. Ramesh, L. T. Florez and V. G. Keramidas, *J. Appl. Phys.* **69**, 4689 (1991).

[33] S. N. Kaul and M. Rosenberg, *Phys. Rev. B* **27**, 5698 (1983).

[34] L. J. Zhu and J. H. Zhao, *Appl. Phys. A*, **111**, 379–387 (2013).

[35] L. J. Zhu, S. H. Nie, K. K. Meng, D. Pan, J. H. Zhao, and H. Z. Zheng, *Adv. Mater.* **24**, 4547 (2012).

[36] See http://www. for details on the Hall resistivity hysteresis and the fits of the data with conventional scaling laws $\rho_{AH}\sim \rho_{xx}^n$ and $\rho_{AH} = a\rho_{xx}+b\rho_{xx}^2$.

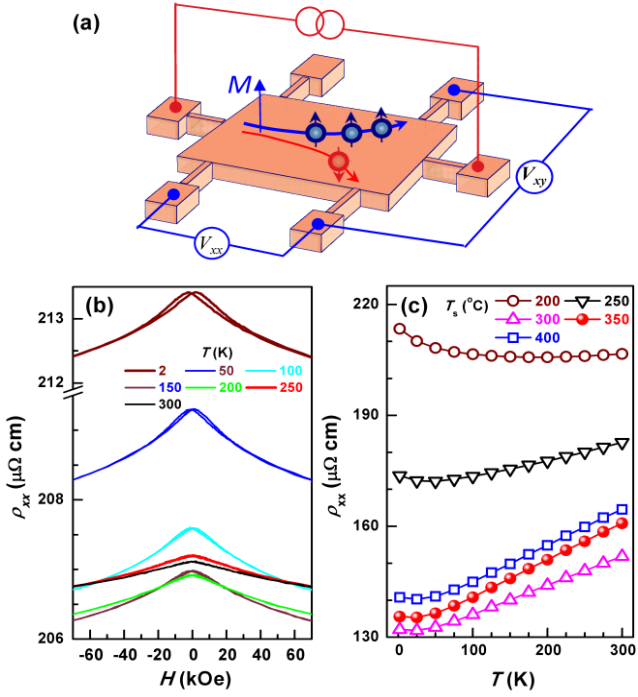

Fig. 1 (a) Schematic of anomalous Hall effect and measurement configuration, (b) $\rho_{xx}$-$H$ curves ($T_s$=200 °C) and (c) $\rho_{xx}$-$T$ curves for $L1_0$-MnAl films with different $T_s$.

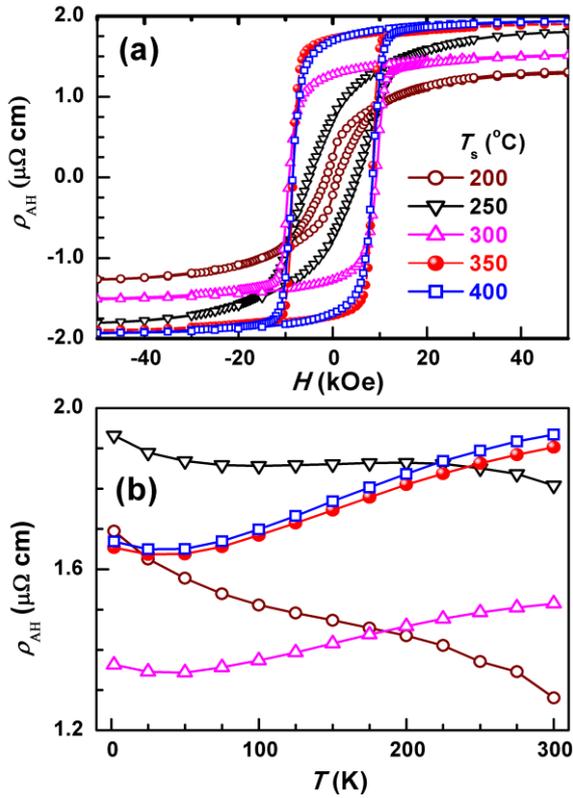

Fig. 2 (a) $H$ (300 K) and (b) $T$ dependence of $\rho_{AH}$ for $L1_0$-MnAl films grown at 200, 250, 300, 350, and 400 °C.

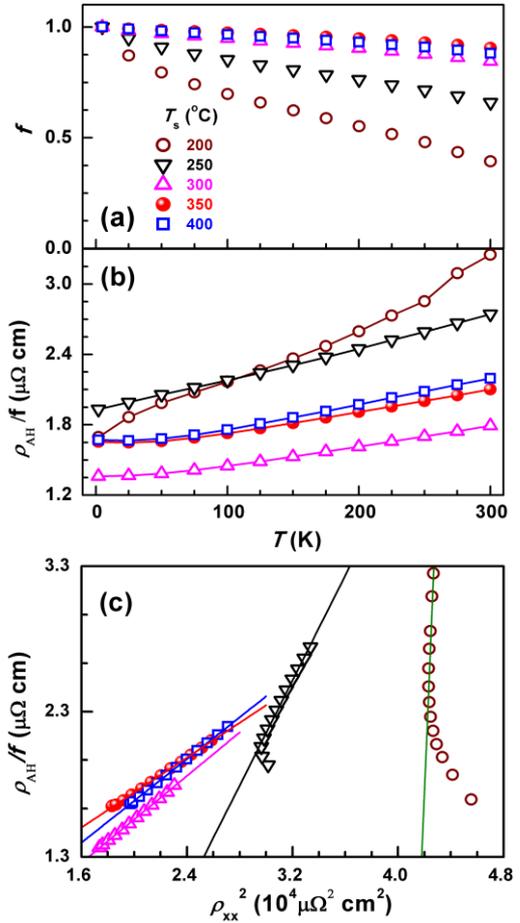

FIG. 3 (a) $f$ versus $T$, (b) $\rho_{AH}/f$ versus $T$, and (c) $\rho_{AH}/f$ versus $\rho_{xx}^2$ for $L1_0$-MnAl films grown at 200, 250, 300, 350, and 400 °C.

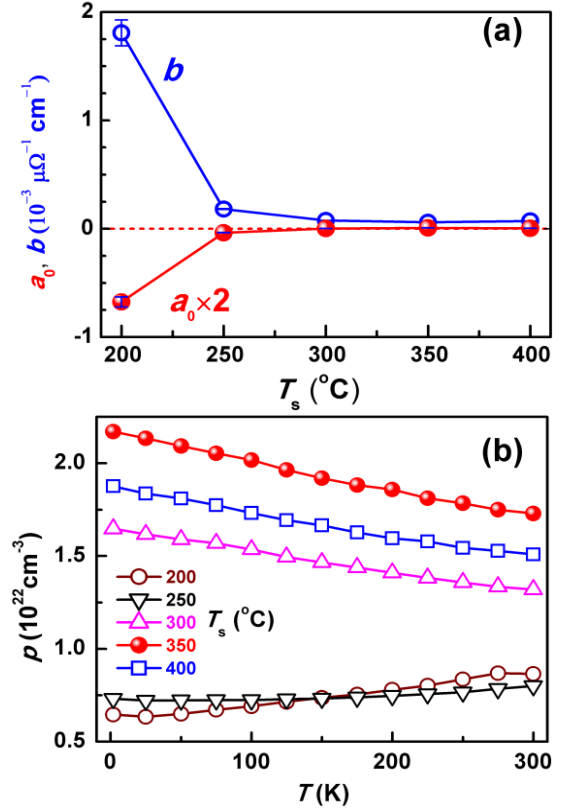

FIG. 4 (a) $a_0$, (b) $b$, and (c) temperature-dependent $p$ for $L1_0$-MnAl films with different $T_s$. For clarity, the values of $a_0$ in (a) is multiplied by a factor of 2.